\documentstyle[psfig]{mn}

\newif\ifAMStwofonts
\AMStwofontstrue

\newcommand{\ds}{\displaystyle}

\newcommand{\be}{\begin{equation}}
\newcommand{\en}{\end{equation}}
\newcommand{\bea}{\begin{eqnarray}}
\newcommand{\ena}{\end{eqnarray}}

\ifoldfss
 \ifCUPmtlplainloaded \else
    \NewTextAlphabet{textbfit} {cmbxti10} {}
    \NewTextAlphabet{textbfss} {cmssbx10} {}
    \NewMathAlphabet{mathbfit} {cmbxti10} {} 
    \NewMathAlphabet{mathbfss} {cmssbx10} {} 
  \fi
  \ifAMStwofonts
    \ifCUPmtlplainloaded \else
      \NewSymbolFont{upmath} {eurm10}
      \NewSymbolFont{AMSa} {msam10}
      \NewMathSymbol{\upi}     {0}{upmath}{19}
      \NewMathSymbol{\umu}     {0}{upmath}{16}
      \NewMathSymbol{\upartial}{0}{upmath}{40}
      \NewMathSymbol{\leqslant}{3}{AMSa}{36}
      \NewMathSymbol{\geqslant}{3}{AMSa}{3E}

      \let\leq=\leqslant 
       
    \fi
  \fi
\fi 

\ifnfssone
  \newmathalphabet{\mathit}
  \addtoversion{normal}{\mathit}{cmr}{m}{it}
  \addtoversion{bold}{\mathit}{cmr}{bx}{it}
  \newmathalphabet{\mathbfit} 
  \addtoversion{normal}{\mathbfit}{cmr}{bx}{it}
  \addtoversion{bold}{\mathbfit}{cmr}{bx}{it}
  \newmathalphabet{\mathbfss} 
  \addtoversion{normal}{\mathbfss}{cmss}{bx}{n}
  \addtoversion{bold}{\mathbfss}{cmss}{bx}{n}
  \ifAMStwofonts
    \ifCUPmtlplainloaded \else
      \UseAMStwoboldmath
      \makeatletter
      \new@mathgroup\upmath@group
      \define@mathgroup\mv@normal\upmath@group{eur}{m}{n}
      \define@mathgroup\mv@bold\upmath@group{eur}{b}{n}
      \edef\UPM{\hexnumber\upmath@group}
      \new@mathgroup\amsa@group
      \define@mathgroup\mv@normal\amsa@group{msa}{m}{n}
      \define@mathgroup\mv@bold\amsa@group{msa}{m}{n}
      \edef\AMSa{\hexnumber\amsa@group}
      \makeatother
      \mathchardef\upi="0\UPM19
      \mathchardef\umu="0\UPM16
      \mathchardef\upartial="0\UPM40
      \mathchardef\leqslant="3\AMSa36
      \mathchardef\geqslant="3\AMSa3E

      \let\leq=\leqslant 

    \fi
  \fi
\fi 

\ifnfsstwo
  \DeclareMathAlphabet{\mathbfit}{OT1}{cmr}{bx}{it}
  \SetMathAlphabet\mathbfit{bold}{OT1}{cmr}{bx}{it}
  \DeclareMathAlphabet{\mathbfss}{OT1}{cmss}{bx}{n}
  \SetMathAlphabet\mathbfss{bold}{OT1}{cmss}{bx}{n}
  \ifAMStwofonts
    \ifCUPmtlplainloaded \else
      \DeclareSymbolFont{UPM}{U}{eur}{m}{n}
      \SetSymbolFont{UPM}{bold}{U}{eur}{b}{n}
      \DeclareSymbolFont{AMSa}{U}{msa}{m}{n}
      \DeclareMathSymbol{\upi}{0}{UPM}{"19}
      \DeclareMathSymbol{\umu}{0}{UPM}{"16}
      \DeclareMathSymbol{\upartial}{0}{UPM}{"40}
      \DeclareMathSymbol{\leqslant}{3}{AMSa}{"36}
      \DeclareMathSymbol{\geqslant}{3}{AMSa}{"3E}

      \let\leq=\leqslant 

    \fi
  \fi
\fi 

\ifCUPmtlplainloaded \else
  \ifAMStwofonts \else 
    \def\upi{\pi}
    \def\umu{\mu}
    \def\upartial{\partial}
  \fi
\fi

\title{Closed universe and the first Doppler-peak of the CMB
spectrum }
\author[S. del Campo]
  {Sergio del Campo$^{1}$\\
$^1$ Instituto de F\'\i sica, Universidad Cat\'olica de Valpara\'\i so, \\
Av Brasil 2950, Valpara\'\i so, Chile. }

\pagerange{\pageref{firstpage}--\pageref{lastpage}} \pubyear{2002}

\begin{document}

\maketitle

\label{firstpage}

\begin{abstract}
We study universe models which  are intrinsically closed and  are
full of a quintessence scalar field, besides the Cold Dark Matter
component. We use these to depict diverse flat Cold Dark Matter
models. With the background geometry specified by the
Friedman-Robertson-Walker metric, we include among them the
standard Cold Dark Matter, the Cosmological Constant Cold Dark
Matter  and the Dark Energy or Quintessence Cold Dark Matter
models. After describing these models, we determine the position
of the first Doppler peak of the Cosmic Microwave Background
anisotropy spectrum for $\Omega_T$ close to one; we also study the
shift parameter $R$.
\end{abstract}


\section{Introduction}
To day, we do not know precisely the exact amount of matter
present in the universe, so that  we ignore its geometry.
Astronomical observations conclude that the matter density related
to baryonic and nonbaryonic cold dark matter is much less than the
critical density value~\cite{Wh-etal}. However, recent
measurements of type Ia distant supernova indicate that in the
universe
there exists an important energy component 
which contributes to a large component of negative pressure, and
thus it accelerates rather than decelerates the
universe~\cite{Pe-etal,Ga-etal}.

Different interpretations have been suggested for explaining the
acceleration of the universe. We distinguish here those related to
the existence of a cosmological constant, and the quintessence or
dark energy models. The former are characterized by a a vacuum
energy density, while the latter are characterized by a scalar
field $\chi$, and its potential $V(\chi)$~\cite{CaDaSt}.

Various tests of cosmological models, including spacetime
geometry, galaxy peculiar velocities, structure formation, and
very early universe descriptions (related to inflation~\cite{Gu})
support a flat universe scenario. Specifically, the
redshift-distant relation for supernova of type Ia, anisotropies
in the cosmic microwave background radiation~\cite{RoHa,Ma-etal}
and gravitational lensing~\cite{Me}  suggest that $ \ds \Omega_T =
1.00 \pm 0.12 \,\,\,\,(\,95\%\,\, cl\,)$~\cite{DB-etal}.

In light of these results, one interesting question to ask is
whether this flatness is due to a sort of compensation among
different components that enter into the dynamical equations. In
the literature we find some descriptions along these lines. For
instance, a closed model with an important matter component with
equation of state given by $P = - \rho / 3$ has been
studied~\cite{Ko}. Here, the universe expands at a constant speed.
Other authors, using the same equation of state, have added a
nonrelativistic component in which the total matter density,
$\Omega_T$, is less than one, thus describing an open
universe~\cite{KaTo}. Also,  flat decelerating universe models
have been simulated~\cite{CrdCHe,CadC,dCCr}. The common fact in
all of these models is that, even though the starting geometry
were other than that corresponding to the critical geometry, all
of these models are indistinguishable from flat models at low
redshift.

In this work we want to describe a closed  universe model composed
of two matter components. One is the usual nonrelativistic dust
matter and the other corresponds to a sort of quintessence-type
matter, designated by the  $Q$ scalar field, which we assume obeys
an equation of state \be \ds P_{_Q} = w_{_Q} \rho_{_Q}\label{eqst}
,\en where, in general, the $w_{_Q}$ parameter is a time dependent
negative. We also assume that its current value is bounded from
the above, $w_{_Q} \leq  -1/3$. The geometry, together with this
matter component, combines in a way such that flat universe
scenarios arise. Among the scenarios which we consider are the
standard Cold Dark Matter or Einstein-de Sitter (sCDM), the
Cosmological Constant Cold Dark Matter ($\Lambda$CDM) and the
Quintessence (or dark energy) Cold Dark Matter ($\chi$CDM) models.
In the latter scenario, a scalar field $\chi$ is added to the
relevant components~\cite{Wa-etal}. In this case, it is assumed
that there exists an equation of state for the dark energy scalar
field $\chi$ given by $P_{\chi} = w_{\chi} \rho_{\chi}$, where
astronomical observations (related to type Ia supernovae
measurements) set an upper limit on the present value of
$w_{_\chi}$, $w_{_\chi} \leq -1/3$~\cite{Ga-etal}.

From the theoretical point of view, anisotropies in the CMB are
related to small perturbations, which are believed to seed the
formation of large-scale structures in the universe. These
anisotropies are sensitive to cosmological parameters such as the
CDM, number of baryons, three-space curvature and the cosmological
constant~\cite{HuSua}. These anisotropies are enhanced by
oscillations of the photon-baryon fluid before decoupling, which
are driven by primordial density fluctuations that depend on the
matter content~\cite{HuSuSi}.

Models with adiabatic and isocurvature fluctuations predict a
sequence of peaks in the power spectrum which are generated by
acoustic oscillations of the photon-baryon fluid at recombination.
The fluctuations, as a function of the wavenumber $k$ go as $cos(k
c_s \tau_{_{LS}} )$ at last scattering, where $c_s$ is the sound
speed and $\tau_{_{LS}}$ is the conformal time at recombination.
In the case of primordial adiabatic fluctuations, these causes a
harmonic series of temperature fluctuation peaks, where $\ds k_m\,
=\,\frac{m\,\pi}{c_s\,\tau_{_{LS}}} $ corresponds to the $m$th
peak. In the isocurvature case, it is found that the acoustic
peaks are $90^0$ degrees out of phase with their adiabatic
counterparts~\cite{HuSub}.

Of particular interest is the height and position of the main
acoustic peak - the so called Doppler peak. This pronounced peak
in the angular power spectrum occurs at multipole $l_{_{LS}}$. The
exact value of $l_{_{LS}}$ depends on both the linear size of the
acoustic horizon and the angular diameter distance from the
observer to the recombination era (last scattering surface). Both
of these quantities are sensitive to a number of cosmological
parameters, essentially to the total density parameter $\Omega_T$,
which is defined to be the ratio between the total matter density
and the critical energy density. In the case in which there is no
contribution from the cosmological constant, it is found that
$l_{_{LS}} \sim
200/\sqrt{\Omega_T}$~\cite{KaSpSu,HuSua,FrNgRo,DoKn,Cr-etal}.

A precise measurement of $l_{_{LS}}$ can efficiently constrain the
density parameter and, specifically, the curvature of the
universe. In fact, in the BOOMERanG (Balloon Observations Of
Millimeter Extragalactic Radiation and Geomagnetic) experiment the
value  $ \ds l_{_{LS}} = (197 \pm 6) \,\,\, (1- \sigma \,\,
error)$ has been reported~\cite{DB-etal}. This, in a model with
$\Lambda = 0$, is consistent with an almost flat geometry, since
this value leads to $\Omega_T \,=\,1.03\, \pm \,0.06$~\cite{RoHa}
when the above expression, $l_{_{LS}} \sim \Omega_T^{-1/2}$, is
used.
For instance, in the case of the $\Lambda$CDM model, we find that
$\Omega_T = \Omega_M + \Omega_{\Lambda}$, in which $ \ds \Omega_M
\equiv \frac{\rho_M^0}{\rho_{_{C}}} = \left ( \frac{8 \pi G}{3
H_0^2} \right )\,\rho_M^0 $ and $ \ds \Omega_{\Lambda}\, \equiv
\left(\frac{\Lambda}{8 \pi G}
\right)\frac{1}{\rho_{_{C}}}\,=\,\frac{\Lambda}{3\,H_0^2}$. Here,
$\rho_M^0$, $\rho_{_{C}}$ and $H_0$ are the present values of the
nonrelativistic matter, the critical densities and the Hubble
constant, respectively, and $G$ is the Newton constant.
$\Omega_{\Lambda}$ and $\Omega_M$ are parameters associated with
the cosmological constant $\Lambda$, and  the matter density is
related to the baryonic and nonbaryonic Cold Dark Matter density,
respectively.  From now on, all quantities with upper (or lower)
zero indexes specify its current values, and we take $c=1$ for the
speed of light.


The paper is presented as follows: In section II we write the
Einstein field equations. In section III we study three specific
models. They are the sCDM, the $\Lambda$CDM, and the $\chi$CDM. In
section IV we proceed to describe the position of the first
Doppler peak for each of these models. Here, we express the first
Doppler peak $l_{_{LS}}$, in terms of the $\Omega_T$ parameter,
considered to be close to one. We also determine the shift
parameter $R$ for the models studied here. We conclude in section
V.


\vspace{-0.25 cm}

\section{\label{sec:level2}The Einstein field equations }

 We start with the effective Einstein action given by
$$ \hspace{-4.0cm}
\ds S\,=\,\int{d^{4}x\,\sqrt{-g}}\,\left
[\,\frac{1}{16\pi\,G}\,{\cal{R}}\,\right .
$$ \be \hspace{2.0cm} \ds \left
.+\,\frac{1}{2}\,(\partial_{\mu}Q)^2\, -\,V(Q)\,+\,{\cal{L}}_{M}
\right ], \label{ac1} \en where ${\cal {R}}$ is the scalar
curvature; $V(Q)$ is the scalar potential associated with the
scalar field $Q$; and ${\cal{L}}_M$ is related to any other matter
component.

We shall assume that the $Q$ field is homogeneous, i.e. it is a
time-dependent quantity only, $Q = Q(t)$; and the spacetime is
isotropic and homogeneous with its metric corresponding to the FRW
metric:
$$ \hspace{-3.0cm} \ds d{s}^{2}= d{t}^{2}- a(t)^{2}\, \left [
\frac{dr^2}{1-k r^2} \right . $$
\begin{equation}
\hspace{2.0cm} \left.  +\,r^2 \left(\,d\theta^2+ sin^2 \theta
\,d\phi^2 \right) \frac{}{}\,\right ], \label{me1}
\end{equation}
where $a(t)$ represents the scale factor, and the  $k$ parameter
takes the values $k\,=\,-\,1,\, 0,\, 1$ corresponding to an open,
flat and closed three-geometry, respectively.  With these
assumptions, action (\ref{ac1}) yields the following field
equations: The time-time component of the Einstein equations
reads: \be \ds H\,^{2}\,=\,\frac{8\pi\,G}{3}\, \left
(\,\rho_{_{M}}\,+\,\rho_{_{Q}}\,\right )\, -\,\frac{k}{a^{2}},
\label{h1} \en
and the evolution equation for the $Q$ scalar field becomes \be
\ds \ddot{Q}\, +\,3\,H\,\dot{Q}\,=-
\,\frac{\partial{V(Q)}}{\partial{Q}}. \label{ddq} \en

Here the overdots denote derivatives with respect to $t$; $ \ds
H\,=\,\frac{\dot{a}}{a}$ defines the Hubble expansion rate;
$\rho_{_{M}}$ and $\rho_{_{Q}}$ are the effective matter energy
density and the average energy density related to $Q$,
respectively.  The Q-energy density is defined by \be \label{rq}
\ds \rho_{_{Q}}\,=\,\frac{1}{2}\dot{Q}^2\,+\,V(Q)\,. \en

We introduce its average pressure $P_{_Q}$ by means of \be
\label{pq} \ds P_{_{Q}}\,=\,\frac{1}{2}\dot{Q}^2\,-\,V(Q)\, . \en
These two quantities are related by the equation of state,
eq.~(\ref{eqst}). By using eqs.~\ref{rq} and~\ref{pq},
eq.~(\ref{ddq}) becomes \be
 \dot{\rho}_{_{Q}}+ 3 H
(\rho_{_{Q}}+P_{_{Q}})=0, \label{erq1} \en
which represents an energy balance for the scalar field $Q$.
Similarly, we have a relation for the nonrelativistic matter
component, i.e. $\dot{\rho}_{_{M}}+ 3 H (\rho_{_{M}}+P_{_{M}})=0$,
which we take to be characterized by the equation of state $P_{_M}
= 0 $, corresponding to a nonrelativistic dust component, in which
case this equation solves to give: $\rho_{_M} \propto a^{-3}$ . In
this way, we have a combination of two noninteracting perfect
fluids: the dust matter component ($\rho_{_{M}}$) and the
quintessence scalar field ($\rho_{_Q}$) component.

Equation (\ref{h1}) may be written as $$ \ds \hspace{-4.0cm} H^2
\,=\,H_0^2\,\left[ \Omega_{_M}\,\left
(\frac{\rho_{_M}}{\rho_{_M}^0}\right ) \,\right. $$ \be
\hspace{3.0cm} \left. +\,\Omega_{_Q}\,\left(
\frac{\rho_{_Q}}{\rho_{_Q}^0}\right)\,+\,\Omega_{_k}\,\left(
\,\frac{a_0}{a} \,\right) ^2 \right].\label{h11}\en  Here, the
present curvature density parameter, $\Omega_k$, and the
quintessence (or dark energy) density parameter, $\Omega_Q$, are
defined by \be \ds \Omega_k \,=\,-\,k\,\left (\frac{1}{a_0\,H_0}
\right )^2,\label{ok}\en and \be \label{oq} \ds \Omega_Q\,
=\,\left (\, \frac{8\,\pi\,G}{3\,H_0^2}\, \right
)\,\rho_{_Q}^0,\en respectively.

In the next section we study the characteristics of the different
models that arise when the nonrelativistic matter component,
$\rho_{_{M}}$ together with equations (\ref{h1}), (\ref{ddq}) and
the equation of state for the scalar field $Q$, are considered, so
that different flat universe models occur. As was mentioned in the
introduction, these models are the sCDM, the $\Lambda$CDM and the
$\chi$CDM.

In order to mimic a flat universe, we  assume that $\rho_{_{Q}}$,
together with the curvature term, combine in a way such that the
following scenarios occur:

$$\hspace{-6.0cm}  \ds  \frac{8 \pi G}{3}\rho_{_{Q}}(t) -\frac{k}{a^2(t)}
$$
\bea \hspace{1.5cm} \equiv \,\left \{
\begin{array}{ll}
   0 & \mbox{for the sCDM mode}, \\
   & \\
   \ds \frac{\Lambda}{3} & \mbox{for the $\Lambda$CDM model}, \\
   & \\
   \ds \frac{8 \pi G}{3}\rho_{_{\chi}}(t) & \mbox{for the $\chi$CDM
   model}.
   \end{array}
   \right.
\label{c1} \ena For future reference we write eq. (\ref{c1}) in
terms of the $\Omega$ parameters:
$$\hspace{-4.0cm}  \ds  \Omega_{_Q}\,\left(
\frac{\rho_{_Q}}{\rho_{_Q}^0}\right)\,+\,\Omega_{_k}\,\left(
\,\frac{a_0}{a} \,\right) ^2
$$
\bea \hspace{1.5cm} \equiv \,\left \{
\begin{array}{ll}
   0 & \mbox{for the sCDM mode}, \\
   & \\
   \ds \Omega_{\Lambda} & \mbox{for the $\Lambda$CDM model}, \\
   & \\
   \ds \Omega_{\chi}\,\left(\,\frac{\rho_{_\chi}}{\rho_{_\chi}^0}\,\right)
   & \mbox{for the $\chi$CDM model},
   \end{array}
   \right.
\label{c2} \ena where, similar to the definitions for
$\Omega_{_M}$ and $\Omega_{_Q}$, we  define $\ds \Omega_{\chi}\,
=\,\left (\, \frac{8\,\pi\,G}{3\,H_0^2}\, \right
)\,\rho_{_{\chi}}^0$ for the present value of the quintessence
density parameter.

Some comments are in order. In the first two cases we could obtain
an explicit expression (as a function of cosmological time) for
the unknown energy density, $\rho_{_Q}$, if we know the scale
factor $a(t)$ as an explicit function of time. On the contrary, in
the third case, we need not only to know the explicit expression
for the scale factor, but also the explicit time dependence of the
dark energy density $\rho_{_\chi}$. Also, since in the first case
the quantity located on the left hand side has to vanish and,
considering that the energy density, $\rho_{_Q}$, can not be
negative, we are forced to consider closed geometries ($k=1$), in
which case $\Omega_k$ become negative.  In the second and third
cases we will also consider the geometry to be closed. In this
case, we could take a complete range for the scale factor $a(t)$,
i. e. $0\leq a(t) < \infty$.


\section{the specific models}

In this section we shall impose the conditions under which a
closed universe ($k = 1$) may look like a flat universe ($k = 0$)
at low redshift. The flat models are characterized by expression
(\ref{c1}) (or equivalently eq.~(\ref{c2})).

\subsection{The Einstein-de Sitter or Standard Cold Dark
Matter (sCDM) Model}

In order to have a closed universe, but one which still has a
nonrelativistic matter density whose value corresponds to that of
a flat universe, we impose the first condition described by
equation~(\ref{c1}), i.e. $ \ds
\rho_{_Q}(a)\,=\,\frac{3}{8\pi\,G\,a^{2}} \label{rq1} $ or,
equivalently, from equation~(\ref{c2}) \be \ds \
\Omega_{_Q}\,\left(
\frac{\rho_{_Q}}{\rho_{_Q}^0}\right)\,=\,-\,\Omega_C\,\left(
\,\frac{a_0}{a} \,\right) ^2,\label{oq11} \en where $\Omega_C$ is
the density parameter for a closed universe, i.e.  $\ds
\Omega_C\,\equiv\,\Omega_{k=1}\,<0$. Note that
equation~(\ref{oq11}) gives at present time $\Omega_Q\, =\,\mid\,
\Omega_C\,\mid$.

When  equation~(\ref{oq11}) is substituted into
equation~(\ref{h11}), the following expression results:  $
H\,^{2}\,=\,8\pi\,G \rho_{M}/3, \label{h2}  $ which gives for a
dust dominated universe: $\ds a(t)=a_0\,(t/t_0)^{2/3}$. Notice
that this expression gives $\Omega_M = 1$ when evaluated at
present time.


Now we are in a position to obtain the intrinsic characteristics
of the scalar field $Q$. From expressions~(\ref{rq})
and~(\ref{pq}), together with the equation of state,
eq.~(\ref{eqst}), we obtain $\ds \dot{Q}(t)\, =\, \sqrt{\left(1 +
w_Q\right)\,\rho_{_{Q}}(t)}$, which gives $ \ds Q\,(t)=\,Q_0\,
(t/t_0)^{\frac{1}{3}} \label{qt} $, where $Q_0$ is defined by $\ds
Q_0=3\,\sqrt{(1+w_Q)}\,(t_0/a_0^2)$.

The same set of equations gives $ \ds
V_Q(t)\,=\,(1-w_Q)\,\rho_{_Q}(t)\,/2=\,V_0\,( t_0/t)^{4/3}
\label{vt} $, where $V_0\,=\,\rho_{_C}\Omega_Q/2$. These solutions
combine in a way such that the scalar potential becomes \be \ds
V_Q(Q)\,=\,V_0\,\left( \frac{Q_0}{Q}\right)^4, \label{vq}\en which
represents a typical potential for a quintessence scalar field.

In order to satisfy the field equation~(\ref{ddq}),  we need to
 take $\ds w_Q \,=\,-\,1/3$. With this value for $w_Q$, the
scalar field $Q$ has properties similar to the matter described in
references~Kolb (1989), Cruz, del Campo \& Herrera (1998), and del
Campo \& Cruz (2000).


\subsection{The Lambda Cold Dark Matter Model ($\Lambda$CDM)}

Following a procedure similar to that of the previous subsection,
we take the second of the three constraints specified by
equation~(\ref{c2}), i.e.
\begin{eqnarray}
\ds \Omega_Q \,\left(\,\frac{\rho_{_Q}(t)}{\rho_{_Q}^0}\,
\right)\,+\,\Omega_C \,\left ( \frac{a_0}{a(t)} \right
)^2\,=\,\Omega_{\Lambda},  \label{c3}
\end{eqnarray}
where the parameters $\Omega_Q$,  $\Omega_C$  and
$\Omega_{\Lambda}$ have already been defined. Equation~(\ref{c3})
evaluated at the present epoch, gives
 $\ds \Omega_Q \,+\,\Omega_C \,=\,\Omega_{\Lambda}$.
Since $\Omega_C\,<\,0$, we must satisfy $\Omega_Q >
\Omega_{\Lambda}$.

Under condition~(\ref{c3}), the time-time component of Einstein
equations becomes analogous to that for a flat universe, where the
usual matter and the cosmological constant form the main matter
components of the model. Thus, equation~(\ref{h11}) reads for a
nonrelativistic perfect fluid: $ \ds H\,^{2}\,=\,H_0^2\,\left [
\,\Omega_{\Lambda}\,+\, \Omega_{M}\,\left( \frac{ a_0}{a}\right)^3
\,\right ]$. Notice that, when this expression is evaluated at
present time, i.e. $t=t_0$, we obtain $ \Omega_M +
\Omega_{\Lambda} \equiv\Omega_T = 1$. For numerical computations
we shall take $\Omega_M = 0.35$ and $\Omega_{\Lambda}= 0.65$. The
latter choice agrees with the amount of cosmological constant,
$\Omega_{\Lambda}\,  < 0.7$, constrained by QSO lensing
surveys~\cite{Koc}.

\begin{figure}
\psfig{file=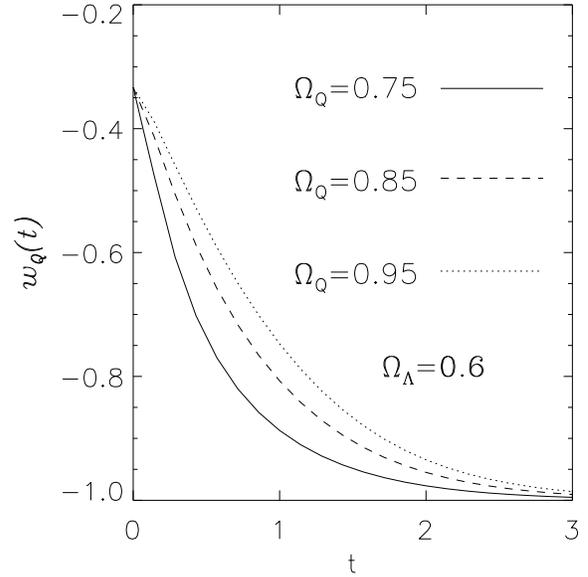,width=3.4in}
\caption{ This graph shows the equation of state
$w_{_Q}^{^{\Lambda CDM}} = P_{_Q}/\rho_{_Q}$ as a function of time
(in units of $H_0$, the present value of the Hubble parameter) for
three different values of the density parameter, $\Omega_Q$ (
$\Omega_Q = 0.75$, $0.85$ and $0.95 $). Here, we have used
$\Omega_\Lambda =0.6$.} \label{EquationOfState}
\end{figure}

Using the definition of the Hubble parameter together with
equation~(\ref{erq1}), we obtain  \be \ds P_{_Q}\,=\,-\,\frac{1}{3
a^2 }\,\frac{d}{d\,a}\,\left( a^3\,\rho_{_Q}\right),\label{pqa}
\en for the effective pressure associated with the $Q$ field. By
substituting eq.~(\ref{c3}) into eq.~(\ref{pqa}), we obtain
\begin{eqnarray}
\label{wt1} \ds w_{_Q}^{^{\Lambda
CDM}}(a)\,=\,-\,\frac{1}{3}\,\left [\frac{
\,(\Omega_Q\,-\,\Omega_\Lambda)\,\left (a_0/a \right
)^2\,+\,3\,\Omega_\Lambda}{\,(\Omega_Q\,-\,\Omega_\Lambda)\,\left
(a_0/a \right )^2\,+\,\Omega_\Lambda}\, \right ],
\end{eqnarray}
for the equation state parameter $w_{_Q}$. Notice that the case
$\Lambda\,=\,0$ gives  $\ds w_{_Q}^{^{\Lambda
CDM}}(a)\,=\,-\,1/3\,=\,$ const., corresponding to the Einstein-de
Sitter model described in the previous subsection. For $\Lambda
\neq 0$, the parameter $w_{_Q}^{^{\Lambda CDM}}(a)$ is always
negative, since in the limit $a \longrightarrow 0$, we get
$w_{_Q}^{^{\Lambda CDM}} \longrightarrow -1/3$ and for $a
\longrightarrow \infty$, we find $w_{_Q}^{^{\Lambda CDM}}
\longrightarrow -1$. Thus, the parameter $w_{_Q}^{^{\Lambda CDM}}$
lies in the range $-1\,<\,w_{_Q}^{^{\Lambda CDM}}\,<\,-1/3$.
\mbox{Figure \ref{EquationOfState}} shows how the equation of
state parameter $w_{_Q}^{^{\Lambda CDM}}$ changes with time, for
three different values of the scalar field density parameter,
$\Omega_Q$; and the parameter $\Omega_\Lambda$ fixed at 0.6.

Another interesting characteristic of the quintessence scalar
field is the form of its scalar potential, $V(Q)$. In order to
determine this form, we consider the definitions~(\ref{rq})
and~(\ref{pq})  together with the equation of state~(\ref{eqst}),
we obtain $$\hspace{-4.0cm} \ds V_Q^{^{\Lambda CDM}}(a)\,=\,V_Q^0
$$
 \be \hspace{2.0cm} \ds \left[ \frac{\,3\,
 \Omega_\Lambda \, + \,2\,\left
(\,\Omega_Q\,-\, \Omega_\Lambda\,\right)\,\left (
\frac{a_0}{a}\,\right )^2}{\,\Omega_\Lambda\, +
\,2\,\Omega_Q}\,\right], \label{vt2} \en where $\ds V_Q^0\,
=\,\frac{1}{3}\,\rho_{_C} \,\left( \,
 \Omega_\Lambda \, + \,2\,\Omega_{_Q}  \right)\equiv \overline{V}_Q\,\left( \,
 \Omega_\Lambda \, + \,2\,\Omega_{_Q}  \right)$
represents the present value of this potential.

On the other hand, from the same equations~(\ref{rq})
and~(\ref{pq}) we get that, after substituting the corresponding
expressions for $\rho_{_Q}$ and $w_{_Q}$ , an explicit expression
for the scalar field $Q$ as a function of the scale factor
$$ \hspace{-4.0cm}
\ds Q(a)\,=\, Q_0\,\left (\frac{a}{a_0}\right)^{1/2}
$$
\be \ds \hspace{2.0cm} \times \,\left[\frac{_2\Large{F}_1\left
(\frac{1}{2},\frac{1}{6};\frac{7}{6};-\left (
\frac{\Omega_\Lambda}{\Omega_M}\right )\left (
\frac{a}{a_0}\right)^3 \right )}{_2\Large{F}_1\left
(\frac{1}{2},\frac{1}{6};\frac{7}{6};-\left (
\frac{\Omega_\Lambda}{\Omega_M}\right ) \right )}\right]
\label{qa1}, \en where $_2F_1$ is the generalized hypergeometric
function and $Q_0$ is defined as $\ds Q_0 = Q(a_0)=
\overline{Q}\,\sqrt{\frac{\Omega_Q -
\Omega_\Lambda}{\Omega_M}}\,_2\Large{F}_1\left
(\frac{1}{2},\frac{1}{6};\frac{7}{6};-\left (
\frac{\Omega_\Lambda}{\Omega_M}\right ) \right )$, with  $\ds
\overline{Q}=\sqrt{\frac{8 \rho_C}{3 H_0^2}}$.

\begin{figure}
\psfig{file=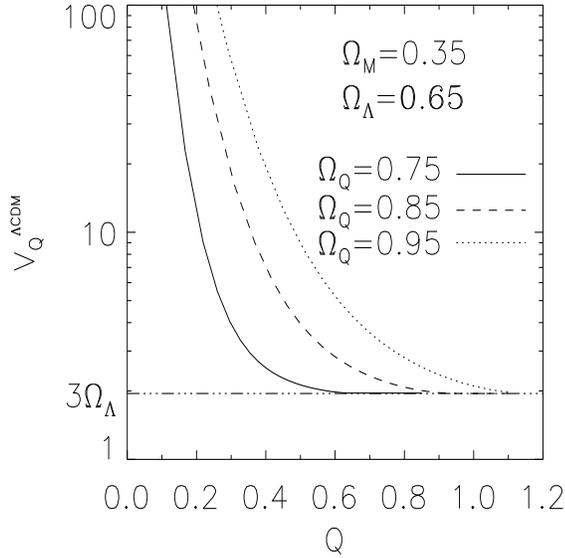,width=3.4in,height=3.4in}
\caption{ Plot of the scalar potential $V_Q$ (in units of
$\bar{V}_Q \equiv \rho_{_C}/3$) as a function of the scalar field
$Q$ (in units of $\bar{Q}\equiv \sqrt{8\rho_{_C}/3H_0^2}\,\,$) for
three different values of the density parameter, $\Omega_Q$ (
$\Omega_Q = 0.75$, $0.85$ and $0.95$). We have taken
$\Omega_M=0.35$ and $\Omega_\Lambda=0.65$.} \label{potential_LCDM}
\end{figure}

By using numerical computations, we can plot the scalar potential
$V_Q$ as a function of the scalar field $Q$. \mbox{Figure
\ref{potential_LCDM}} shows the plot for three different values of
the parameter density $\Omega_Q$, ( $\Omega_Q = 0.75$, $0.85$ and
$0.95$). The other parameters, $\Omega_M$ and $\Omega_\Lambda$ are
fixed at values 0.35 and 0.65, respectively.
Note that at sufficiently higher values of $Q$ the potential
approaches a constant value given by
$3\overline{V}_Q\,\Omega_\Lambda$. This value becomes independent
of the parameter $\Omega_Q$.


\subsection{The Quintessence (or Dark Energy) Cold Dark
Matter Model ($\chi$CDM)}

In this case we consider the following constraint equation: \be
\ds \Omega_{_Q}\,\left(
\frac{\rho_{_Q}}{\rho_{_Q}^0}\right)\,+\,\Omega_{_c}\,\left(
\,\frac{a_0}{a} \,\right) ^2 =
\Omega_{\chi}\,\left(\,\frac{\rho_{_\chi}}{\rho_{_\chi}^0}\,\right),
\label{c4}\en which reduces the time-time component of Einstein
equations to \be \ds H\,^{2}\,=\,H_0^2\,\left [ \,
\Omega_{M}\,\left( \frac{ a_{_0}}{a}\right)^3 +\,
\Omega_{\chi}\,\left( \frac{ \rho_{_\chi}}{\rho_{_\chi}^0}\right)
\,\right ], \label{h4}\en where, just as before, we have
considered dust to be the regular matter, $\rho_{_M}$.

These two latter equations together with the evolution equations
for the scalar fields $\chi$ and $Q$ form the basic set of
equations for our model. In order to solve this set of equations,
we need to introduce the equations of state for the quintessence
scalar field components $Q$ and $\chi$. We assume that the $\chi$
field component is characterized by a constant equation state
parameter that lies in the range $-1< w_{_{\chi}} < - 0.6$.
Instead, we consider $w_{_Q}$ to be a variable quantity whose
actual value lies in the same range as $w_{_Q}$.

We can use the definition of  $P_{_{\chi}}$ and $\rho_{_{\chi}}$
in terms of the scalar field $\chi$, together with the equation of
state that relates these quantities, for obtaining $\chi$ field as
a function of the scale factor $a$. The result is
$$ \ds \hspace{-5.0cm}\chi(a)\,=\,\chi_{_{0}}\left(\frac{a}{a_0}\right)^{-3w_\chi/2} $$
\be \hspace{1.0cm} \ds \times
\left[\,\frac{_2F_1\left(\frac{1}{2}, \frac{1}{2};
\frac{3}{2};-\left(\frac{\Omega_{\chi}}{\Omega_M}\right)\,\left(\frac{a}{a_0}
\right)^{- 3\, w_{_{\chi}}} \right)}{_2F_1\left(\frac{1}{2},
\frac{1}{2};
\frac{3}{2};-\left(\frac{\Omega_{\chi}}{\Omega_M}\right)
\right)}\right], \label{chia} \en where $\chi_{_{0}}$ is given by
$\ds \chi_{_{0}}=\widetilde{\chi}\,\,_2F_1\left(\frac{1}{2},
\frac{1}{2};
\frac{3}{2};-\left(\frac{\Omega_{\chi}}{\Omega_M}\right) \right)$,
with $\widetilde{\chi}=
\sqrt{4\rho_c/9H_0^2}\sqrt{\Omega_\chi(1+w_\chi)/\Omega_M
w_\chi^2}$.

In a similar way we obtain  \be \ds
V_{_{\chi}}(a)\,=\,V_\chi^0\,\left( \frac{a_0}{a}\right)
^{3(1+w_{_{\chi}})}, \label{vchi} \en for the scalar potential
$V_{_{\chi}}$ as a function of the scale factor $a$, where the
present value of this potential is given by $\ds
V_\chi^0\,=\,\frac{1}{2}(1-
w_{_{\chi}})\,\rho_{_{C}}\Omega_{\chi}$.

\begin{figure}
\psfig{file=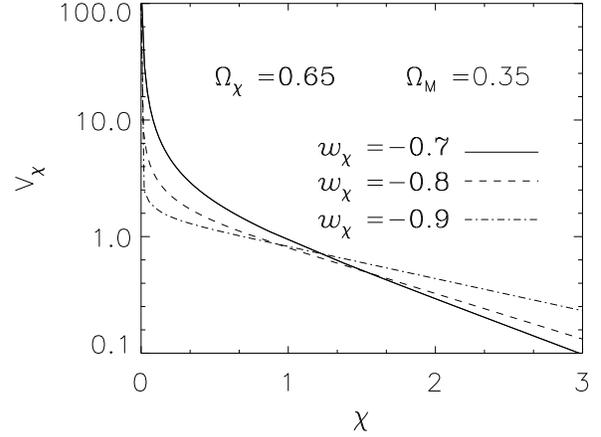,width=3.4in}
\caption{ This graph shows the scalar potential $V_{_{\chi}}$ (in
units of $\rho_{_{C}}/2$) as a function of the scalar field $\chi$
(in unit of $\sqrt{4\rho_{_{C}}/9 H_0^2}$) for three different
values of the equation state parameter, $w_{_{\chi}}=-0.7$,$ -0.8$
and $ -0.9 $. Here, we have taken $\Omega_M=0.35$ and $\Omega_\chi
=0.65$.} \label{V(chi)}
\end{figure}

\mbox {Figure \ref{V(chi)}} shows the plot of the scalar potential
$V_{_{\chi}}$ as a function of the scalar field $\chi$, for
different values of the state equation parameter $w_{_{\chi}}$;
and the parameters $\Omega_\chi$ and $\Omega_M$ have been fixed at
0.65 and 0.35, respectively. This form of potential has been
described in the literature~\cite{DeRaSaSt}.
Note that, as long as $w_{_{\chi}} \longrightarrow -1$, the
potential $V_{\chi} \longrightarrow const. \equiv
\rho_{_C}\,\Omega_{\chi}$, i.e., the model becomes equivalent to
the $\Lambda$CDM.

\begin{figure}
\psfig{file=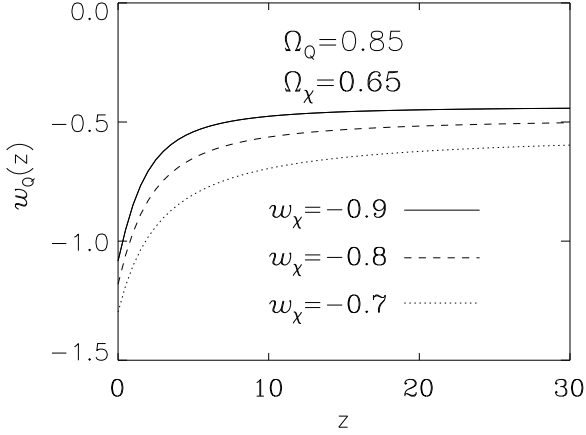,width=3.4in}
\caption{This plot shows    $w_Q$ (in units of $|w_{_{Q}}^0|$) as
a function of the redshift $z$ for three different values of the
equation state parameter, $w_{_{\chi}}= -0.7, -0.8, -0.9 $. We
take $\Omega_Q=0.85$ and $\Omega_\chi=0.65$.} \label{wz_QCDM}
\end{figure}

Following an approach analogous to that done in the previous
subsection, we find that $w_Q$ is given by
$$ \hspace{-2.0cm}w_{_Q}^{^{\chi CDM}}(a)=-\mid w_{_Q}^0 \mid \left(
\frac{1+\beta}{1-3\beta\, w_{_{\chi}}}\right) $$
 \be \ds \hspace{1.0cm}\times \left[ \frac{1-3\beta\,
w_{_{\chi}} \left( \frac{a}{a_0}\right)^{-3 w_{_{\chi}}
-1}}{1+\beta\left( \frac{a}{a_0}\right)^{-3
w_{_{\chi}}-1}}\right],\en where $\beta = \Omega_\chi/(\Omega_Q -
\Omega_\chi)$ and $w_{_Q}^0$ is the present value of $w_{_Q}(a)$
defined by $w_{_Q}^0=P_Q^0/\rho_{_{Q}}^0$.
\mbox{Figure~\ref{wz_QCDM}} shows its dependence on the redshift
$z$ defined as $z\equiv a_0/a-1$, for three different values of
$w_{_{\chi}}$. As before, we have chosen $\Omega_Q=0.85$ and
$\Omega_\chi=0.65$.

In this case the scalar field $Q$ becomes given by \be \ds
Q(a)=\overline{Q}\int_0^{a/a_0}\sqrt{
\frac{1+\frac{3}{2}\beta(1+w_\chi)x^{-(1+3w_\chi)}}{x+\frac{\Omega_\chi}{\Omega_M}
x^{-(1+3w_\chi)}}} dx,\label{Q_QCDM}\en where
$\overline{Q}=\sqrt{\frac{1}{4 \pi
G}\left(\frac{\Omega_Q-\Omega_\chi}{\Omega_M} \right)}$.

On the other hand, the scalar potential $V_Q^{^{\chi CDM}}(a)$ is
given by
$$\ds \hspace{-5.0cm}V_Q^{^{\chi CDM}}(a) = V_Q^0\left(\frac{a_0}{a}\right)^2$$
\be \ds \hspace{2.0cm} \times \left[
\frac{4+3\beta(1-w_\chi)\left(\frac{a}{a_0}\right)^{-(1+3w_\chi)}}{4+3\beta(1-w_\chi)}
\right], \label{Vq_QCDM}\en where $V_Q^0 = \frac{\rho_C
\Omega_Q}{2(1-w_\chi)}\left[4/3 + \beta(1-w_\chi)\right]$.

\begin{figure}
\psfig{file=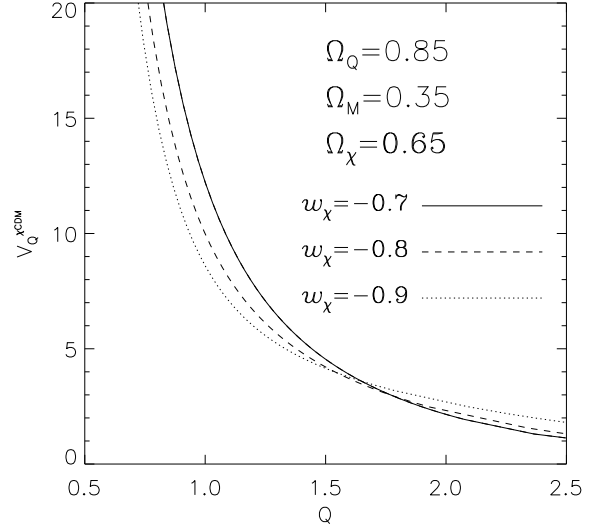,width=3.4in}
\caption{This plot shows $V_Q$ (in units of $\rho_{_C}/2$) as a
function of the scalar field $Q$ (in units of $1/\sqrt{4 \pi G}$)
for three different values of the equation state parameter,
$w_{_{\chi}}= -0.7, -0.8, -0.9 $. Here, we have taken the values
$\Omega_Q=0.85$, $\Omega_\chi=0.65$ and $\Omega_M=0.35$.}
\label{potencial2_QCDM}
\end{figure}

\mbox {Figure \ref{potencial2_QCDM}} shows the scalar potential
$V_Q^{\chi CDM}$ (in units of $\rho_{_C}/2$) as a function of the
scalar field $Q$ (in units of $1/\sqrt{4 \pi G}$) for three
different values of the parameter $w_\chi$ ($w_{_{\chi}}= -0.7,
-0.8, -0.9 $). Here, we have taken $\Omega_Q=0.85$,
$\Omega_M=0.35$ and $\Omega_\chi=0.65$.

Note that this scalar potential decreases when $Q$ increases. This
potential asymptotically tends to vanishing for $a \longrightarrow
\infty $. This implies that, asymptotically, the effective
equation of state becomes  $P_Q = \rho_{_{Q}}$ (for $\dot{Q}\,\neq
0$), corresponding to a stiff fluid.

\section{The first Doppler peak of the CMB spectrum}

In this section, we are going to describe the position of the
first Doppler peak ($l_{_{LS}}$) for the different models studied
in the previous section.

The scales, which are important in determining the shape of the
CMB anisotropy spectrum are the sound horizon $d_s$ at the time of
recombination, and the angular diameter distance $d_A$ to the last
scattering surface. The former defines the physical scales for the
Doppler peak structure that depends on the physical matter density
($\Omega_M$), but not on the value of the cosmological constant
($\Omega_\Lambda$) or spatial curvature ($\Omega_C$), since these
are dynamically negligible at the time of recombination~\cite{EB}.
The latter depends practically on all of the parameters and is
given by (for a closed universe): \be \ds d_A = \frac{1}{
H_0(1+z_{_{LS}})}\frac{1}{\sqrt{\mid\Omega_C \mid}}sin
\left(\sqrt{\mid\Omega_C \mid}y_{_{LS}} \right),\label{da}\en with
$y_{_{LS}}$ determined from \be\ds y_{_{LS}}=
H_0\,\int_{x_{_{LS}}}^1\frac{ dx}{x^2 H(x) },\label{yls}\en where
$x=a/a_0$ and $H(x)$ is the Hubble parameter obtained from the
time-time component of the Einstein equations. We may write for
the localization of the first Doppler peak
$$ \ds l_{_{LS}} \propto \frac{d_A}{d_S},$$ where the constant of
proportionality depends on both the shape of the primordial power
spectrum and the Doppler peak number~\cite{HW}. Since we are going
to keep the $\Omega_M$ parameter fixed, we shall take $l_{_{LS}}
\approx d_A$, up to a factor that depends on $\Omega_M$ and
$z_{_{LS}}$ only.

For the $sCDM$ model we find that $$ \ds \hspace{-3.5cm}
l_{_{LS}}^{sCDM}\, \approx \,\frac{1}{H_0
(1+z_{_{LS}})}\,\frac{1}{\sqrt{\Omega_M}}
$$
\be \ds \hspace{2.0cm} \times\,\sin
\left[2\,\sqrt{\frac{\Omega_Q}{\Omega_M}}\left(
1-\frac{1}{\sqrt{1+z_{_{LS}}}}\right) \right]. \label{Scdm}\en
Note that, for $\Omega_Q \ll \Omega_M$ (which means  that the
curvature term is very small), we find that
$l_{_{LS}}^{sCDM}\,\sim \,1/\sqrt{\Omega_M}=1/\sqrt{\Omega_T}$
which  agrees with the result obtained by Frampton {\it et
al}~\cite{FrNgRo} for $\Omega_\Lambda=0$.

In the $\Lambda$CDM model, we find that
$$ \ds \hspace{-3.0cm}
l_{_{LS}}^{^{\Lambda CDM}} \, \approx \,\frac{1}{H_0
(1+z_{_{LS}})}\,\frac{1}{\sqrt{\Omega_Q\,-\,\Omega_\Lambda}}
$$
\be \ds \hspace{0.7cm} \times\,\sin
\left[\,\sqrt{\Omega_Q\,-\,\Omega_\Lambda}
\int_0^1\,\frac{dx}{\sqrt{\Omega_M\,
x\,+\,\Omega_\Lambda\,x^4}}\,\right], \label{Lcdm}\en where we
have set the lower limit on the integral equal to zero, since
$z_{_{LS}} \gg 1$, and therefore we may take $x_{_{LS}} \equiv
1/(1+z_{_{LS}})\approx 0$.

From the equation~$\ds \Omega_Q \,+\,\Omega_C
\,=\,\Omega_{\Lambda}$ together with the expression~$\Omega_M +
\Omega_\Lambda =\Omega_T=1$, we may write
$\Omega_C=1-\Omega_M-\Omega_Q$. Now, following Wienberg~\cite{We},
we will keep fixed the $\Omega_M$ parameter with $\Omega_T$ close
to one and $\Omega_Q$ close to $\Omega_\Lambda$. Thus, we find
that \be \ds l_{_{LS}}^{^{\Lambda CDM}}  \sim
\Omega_T^{-\eta},\label{lcdm1}\en where \be \ds
\eta=\left(\frac{\partial \ln l_{_{LS}}^{^{\Lambda CDM}}}{\partial
\mid \Omega_C \mid}\right)_{\Omega_\Lambda = 1-\Omega_M}=
\frac{1}{6}{\cal I}_1^2-\frac{1}{2}\frac{{\cal I}_2}{{\cal
I}_1}\label{eta},\en with \be \ds {\cal I}_1\equiv
\int_0^1\frac{dx}{\left[(1-\Omega_M)x^4+\Omega_M x\right]^{1/2}},
\en and \be \ds {\cal I}_2\equiv
\int_0^1\frac{x^4\,dx}{\left[(1-\Omega_M)x^4+\Omega_M
x\right]^{3/2}}.\label{Int1}\en These expressions yield $\eta =
2.45$ for $\Omega_M=0.2$ and $\eta = 11/18$ for $\Omega_M=1.0$.
The latter value should be compared with that corresponding to the
$\Omega_{\Lambda} = 0$ case, in which $\eta = 1/2$.

\begin{table*}
\caption{\label{table}This table shows the exponent parameters
$\eta$ of Eq.~(\ref{lcdm1}) for the $\Lambda$CDM and the $\chi$CDM
models, where we have used different values of the parameter
$\Omega_M$. For the latter model we have taken the values -0.7,
-0.8 and -0.9 for the parameter $w_{\chi}$. }
\begin{tabular}{ccccc}
 &   & \multicolumn{3}{c}{ } \\ \hline
 $\Omega_M$&   $\eta^{\Lambda CDM}$& &$\eta^{\chi CDM}$ & \\
 \hline
 & &$w_{\chi}=-0.7$ &$w_{\chi}=-0.8$ & $w_{\chi}=-0.9$ \\ \hline
 0.2& 2.422 &2.152&2.258& 2.346\\
 0.3& 1.727&1.571&1.633& 1.684\\
 0.4& 1.350 &1.248& 1.289 &1.322\\
 & & & & \\
 1.0& 0.595&0.571 &0.581& 0.588\\
\end{tabular}
\end{table*}

For the $\chi CDM$ model we find that
$$ \ds \hspace{-3.0cm}
l_{_{LS}}^{^{\chi CDM}} \, \approx \,\frac{1}{H_0
(1+z_{_{LS}})}\,\frac{1}{\sqrt{\Omega_Q\,-\,\Omega_\chi}}
$$
\be \ds \hspace{0.7cm} \times\,\sin
\left[\,\sqrt{\Omega_Q\,-\,\Omega_\chi}
\int_0^1\,\frac{dx}{\sqrt{\Omega_M\,
x\,+\,\Omega_\chi,x^{1-3w_\chi}}}\,\right], \label{xcdm}\en
Following an approach similar to that described above, we now use
$\Omega_Q\,+\,\Omega_C\,=\,\Omega_{\chi}$ together with
$\Omega_M+\Omega_\chi\equiv \Omega_T=1$ and, taking $\Omega_Q$
close to $\Omega_\chi$, we find  an expression similar to
eq.~(\ref{lcdm1}), with $\eta$ given by eq.~(\ref{eta}), but now
the integrals are given by \be \ds \widetilde{{\cal I}}_1\equiv
\int_0^1\frac{dx} {\left[(1-\Omega_M)x^{1-3w_\chi}+\Omega_M
x\right]^{1/2}}, \en and \be \ds \widetilde{{\cal I}}_2\equiv
\int_0^1\frac{x^{1-3w_\chi}\,dx}{\left[(1-\Omega_M)x^{1-3
w_\chi}+\Omega_M x\right]^{3/2}}.\label{Int2}\en

The values of the exponent $\eta$ appearing in
equation~(\ref{lcdm1}) are tabulated in table~\ref{table} for the
different models treated here.

One important parameter that describes  the dependence of the
first Doppler peak position on the different parameters that
characterize any model is the shift parameter $R$. This parameter
is related to the geometry of the universe and, for closed models,
it may be defined as~\cite{EB,MG,M} \be \ds R =
\sqrt{\frac{\Omega_M}{|\Omega_C|}}\sin\left[\sqrt{|\Omega_C|}\,\,
y_{LS} \right],\en where $y_{LS}$ is defined by
equation~(\ref{yls}).

For the sCDM model, this parameter is given by
$$\ds R_{sCDM} = \sqrt{\frac{\Omega_M}{\Omega_Q}}~
sin\left[2\sqrt{\frac{\Omega_Q}{\Omega_M}}~\left(1-\frac{1}{1+z_{LS}}
\right)\right],$$ which, at first glance, seems to be a quantity
that depends on $\Omega_Q$ and $\Omega_M$ parameters. But we know
that, in this model, the parameter $\Omega_M$ gets the value one.
Here, we have used the equality $|\Omega_C|= \Omega_Q$.

For the $\Lambda$CDM model we find that the parameter R is given
by
$$ \ds \hspace{-2.5cm} R_{\Lambda CDM}
=\sqrt{\frac{\Omega_M}{\Omega_Q-\Omega_{\Lambda}}}~
\sin\left[\sqrt{\Omega_Q-\Omega_{\Lambda}}\right.$$ \be
\hspace{3.5cm} \ds \left. \times \int_0^1\frac{dx}{\sqrt{\Omega_M
x + \Omega_{\Lambda} x^4}} \right] .\en Here, we have used the
relation $|\Omega_C| =\Omega_Q - \Omega_{\Lambda}$ (with $\Omega_Q
> \Omega_{\Lambda}$) and we have considered $x_{LS}\approx 0$.

\begin{figure}
\psfig{file=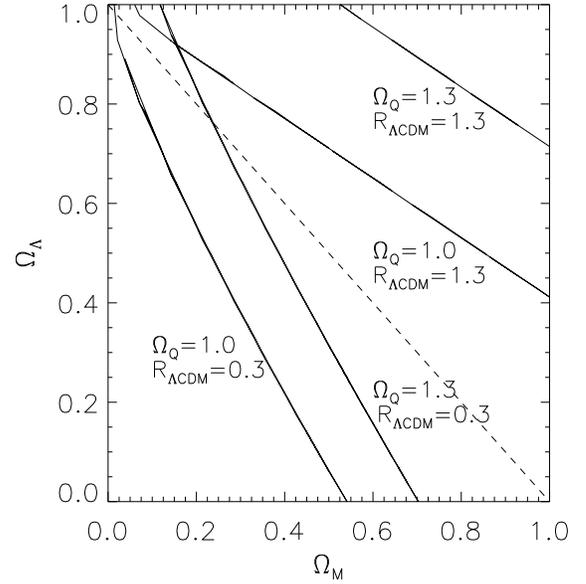,width=3.4in}
\caption{This plot shows a set of lines $R=$ const. for different
values of the parameter $\Omega_Q$. The values of $R$ and
$\Omega_Q$ are given next to each line. We have also included in
this plot the line that joins the points $(1.0;0.0)$ and
$(0.0;1.0$ (dashed line).} \label{RLCDM}
\end{figure}

In \mbox {Figure \ref{RLCDM}} we have plotted a set of lines $R=$
const. for different values of the parameter $\Omega_Q$. The
values of $R$ and $\Omega_Q$ are given next to each line. Notice
that, when we increase the value of $\Omega_Q$ (by keeping the
value of $R$ fixed), the $R=$ const. lines are moved towards
greater values of the $\Omega_M$ parameter. This means that the
Doppler peaks are shifted towards greater angular scale
values~\cite{MG}.

For the $\chi$CDM model we find that the $R$ paremeter becomes
given by the following expression: $$\hspace{-2.5cm} \ds R_{\chi
CDM}= \sqrt{\frac{\Omega_M}{\Omega_Q-\Omega_{\chi}}}~
\sin\left[\sqrt{\Omega_Q-\Omega_{\chi}}\right.
$$
\be \hspace{3.5cm} \ds \left. \times \int_0^1\frac{dx}{\sqrt{\Omega_M x +
\Omega_{\chi} x^{1-3w_{\chi}}}} \right] \label{R3}.\en

\begin{figure}
\psfig{file=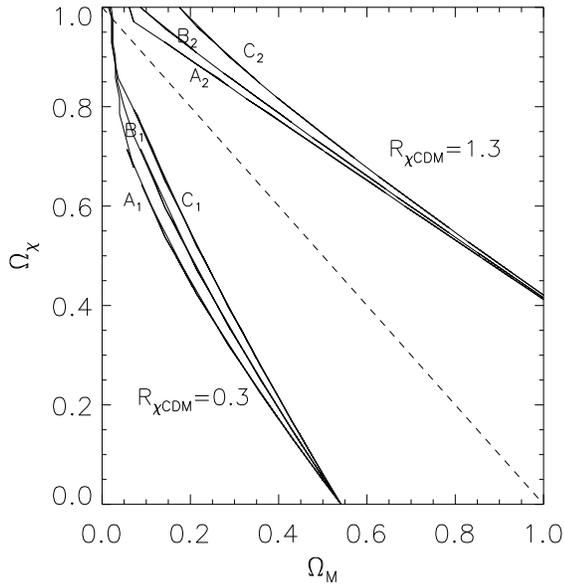,width=3.4in}
\caption{This plot shows two sets of lines of constant $R$ in the
$\Omega_{\chi}$ - $\Omega_M$ plane for three different values of
the state equation parameter; $w_{\chi} = -0.3$ (A$_1$ and A$_2$
lines), $w_{\chi} = -0.5$ (B$_1$ and B$_2$ lines) and $w_{\chi} =
-0.9$ (C$_1$ and C$_2$ lines). The value of $R$ is given next to
each set of lines.} \label{ChiCDM}
\end{figure}

In \mbox {Figure \ref{ChiCDM}}~we have plotted two sets of
$R_{\chi CDM} = const.$ contours for three different values of
$w_{\chi}$ ($w_{\chi}=-0.3, -0.5, -0.9$). In these curves we have
kept fixed $\Omega_Q$ equal to one. Notice that, for small values
of the $\Omega_{\chi}$, these curves almost overlaps.

We should note that the sCDM and $\Lambda$CDM models are special
cases of the $\chi$CDM model. They are obtained from
equation~(\ref{R3}) by taking $\Omega_{\chi} =0$ and
$\Omega_{\chi} = \Omega_{\Lambda}$ with $w_{\chi}=-1$,
respectively. At this point, we should add that more precise
future astronomical measurements of the location of the first
Doppler peak (and its corresponding characteristics) will
certainly supply information on which of these models (or another
one) is more appropriate for describing the universe we live in.

\section{conclusions}

In this paper we have described closed universe models in which,
apart from the usual Cold Dark Matter component, we have included
a quintessence scalar field $Q$. We have fine tuned the
quintessence component together with the curvature term for
getting a flat model in which three different models were
described. These models were the standard Cold Dark Matter ($sCDM$
model, characterized by $\Omega_T \equiv 1$), the Cosmological
constant Cold Dark Matter ($\Lambda CDM$ model, characterized by
$\Omega_T = \Omega_M+\Omega_\Lambda \equiv 1$) and the
Quintessence (or Dark Energy) Cold Dark Matter ($\chi CDM$, model
characterized by $\Omega_T= \Omega_M+\Omega_\chi \equiv 1$). In
all of them we have described the properties of the scalar field
$Q$. The characterization of the scalar field $Q$ in the different
models comes from the determination of the  scalar potential
$V(Q)$. In all of these models, this potential decreases when the
scalar field $Q$ increases. This property seems to be common to
all of the dark energy potentials. In the $\Lambda CDM$ model,
this potential approaches a constant given by $3\overline{V}_Q
\Omega_\Lambda$ at a large scale factor. In the other two models,
the potential $V(Q)$ goes to zero, asymptotically. Since all of
these models are indistinguishable from flat models at enough low
redshift (say $z \sim 1$), we expect that with an appropriate fine
tuning, it will be possible to consolidate the supernova
measurements with closed universe models.

As an applicability of the different models described above, we
have determined the position of the first Doppler peak together
with the shift parameter $R$. For $\mid \Omega_C\mid \approx 0$,
we have found that the first Doppler peak is quite sensitive to
the mass density values ($\Omega_M$).

For a fixed $\Omega_M$ at 0.4 and $\Omega_T \approx 1$, the first
Doppler peak behaves as $\Omega_T^{-\eta}$, with the exponent
$\eta$ given by 0.500 (for the $sCDM$ model), 1.350 (for the
$\Lambda CDM$ model), 1.248 (for the $\chi CDM$ model with $w_\chi
=-0.7$), 1.289 (for the $\chi CDM$ model with $w_\chi =-0.8$) and
1.322 (for the $\chi CDM$ model with $w_\chi =-0.9$). We may
compare these values with that specified by Weinberg, which
results to be 1.244. This value agrees with that obtained from the
$\chi CDM$ model, in which $w_\chi =-0.7$. Some values of the
$\eta$ exponent have been tabulated in table~\ref{table}. In this
table we have excluded the value corresponding to the $sCDM$
model, where $\eta = 1/2$ is found. Precise measurements of the
location of the first Doppler peak (together with the other peaks
and their properties) can supply information on the parameters
$\Omega_{\Lambda}$ (or $\Omega_\chi$) and $\Omega_T$, from which
we could obtain an appropriate value for the parameter $\Omega_Q$.

Secondly, we have determined the shift parameter $R$. In the
$\Lambda$CDM model, this parameter is highly sensitive to the
value of $\Omega_Q$. Something similar occurs in the $\chi$CDM
model. We have plotted curves $R$= const. in the graph
$\Omega_{\chi}$ v/s $\Omega_M$ for different values for the
$w_{\chi}$ parameter. For $\Omega_{\chi} \approx 1$ the $R$=const.
curves became separated. But, for $\Omega_{\chi} \approx 0$, they
began to come together. It is interesting to consider the value $R
= 1$, since this case allows us to determine a precise value for
the $\Omega_Q$ parameter. For the sCDM model we obtain that
$\Omega_Q = 0$; meanwhile, for the other two cases, $\Omega_Q =
\Omega_{\Lambda}$ and $\Omega_Q = \Omega_{\chi}$ for the
$\Lambda$CDM and $\chi$CDM models, respectively. We may conclude
that, as far as we are concerned with the observed acceleration
detected in the universe and the location of the first Doppler
peak, we will be able to utilize a closed model to describe the
universe we live in.

\section*{Acknowledgments}

I am grateful to Francisco Vera for plotting assistance and Paul
Minning for carefully reading the manuscript. Discussion with M.
Cataldo, N. Cruz, S. Lepe, F. Pe\~na and P. Salgado are
acknowledged. I also acknowledge the Universidad de Concepci\'on
and Universidad del Bio-Bio for partial support of the Dichato
Cosmological Meeting, where part of this work was done. This work
was supported from COMISION NACIONAL DE CIENCIAS Y TECNOLOGIA
through FONDECYT Projects N$^0$ 1000305 and 1010485 grants, and
from UCV-DGIP N$^0$ 123.752.

\label{lastpage}

\end{document}